\documentclass[showkeys]{revtex4}

\usepackage{graphicx}
\usepackage{amsmath,amsthm,amssymb,bm,amsfonts}

\def\qbar{{\overline{q}}}

\def\qbold{{\mathbf{q}}}

\def\eto{{\rm{e}}}

\def\Dtilde{\tilde{D}}
\def\gammatilde{\tilde{\gamma}}

\def\dbar{{\bar{d}}}

\newcommand{\be}{\begin{eqnarray}}
\newcommand{\ee}{\end{eqnarray}}

\begin{document}

\title{\bf Electroweak  boson production in double parton scattering}

\author{Krzysztof Golec-Biernat}\email{golec@ifj.edu.pl}
\affiliation{Institute of Nuclear Physics Polish Academy of Sciences, 31-342 Cracow, Poland}
\affiliation{Faculty of Mathematics and Natural Sciences, University of Rzesz\'ow,  35-959 Rzesz\'ow, Poland}

\author{Emilia Lewandowska}\email{emilia.lewandowska@ifj.edu.pl}
\affiliation{Institute of Nuclear Physics Polish Academy of Sciences, 31-342 Cracow, Poland}

\begin{abstract}
We study the $W^+W^-$ and $Z^0Z^0$ electroweak boson production in double parton scattering
using QCD evolution equations for double parton distributions. In particular, we
analyze the impact  of  splitting terms in the evolution equations on the 
double parton scattering cross sections. Unlike the standard terms, the splitting terms 
are not suppressed for large values of the relative momentum of two partons in
the double parton scattering.  Thus, they play an important role which we discuss in detail
for the single splitting contribution to the cross sections under the study.

\end{abstract}

\keywords{quantum chromodynamics, double parton scattering,  double parton distributions
evolution equations, electroweak bosons}

\maketitle

\section{Introduction}

The double parton scattering (DPS) in high-energy  hadron scattering is a process in which 
two hard interactions with large scales (much bigger that nucleon mass) take place in one scattering event.
Such a process is usually interpreted in QCD as scattering of two pairs of partons (quarks or gluons) from
incoming hadrons. 
The DPS is the simplest  multiparton process with hard scales which allows one to gain information
on parton correlations inside hadrons. Thus, it has been studied  for many years 
from both theoretical \cite{Shelest:1982dg,Zinovev:1982be,Ellis:1982cd,Bukhvostov:1985rn,
Snigirev:2003cq,Korotkikh:2004bz,Gaunt:2009re,Blok:2010ge,Ceccopieri:2010kg,Diehl:2011tt,Gaunt:2011xd,Ryskin:2011kk,
Bartels:2011qi,Blok:2011bu,Diehl:2011yj,
Manohar:2012jr,Ryskin:2012qx,Gaunt:2012dd,Snigirev:2014eua} 
and phenomenological sides \cite{Sjostrand:1987su,DelFabbro:1999tf,Kulesza:1999zh,DelFabbro:2002pw,Cattaruzza:2005nu,Sjostrand:2004pf,Berger:2009cm,Gaunt:2010pi,Snigirev:2010tk,Kom:2011bd,Berger:2011ep,Kom:2011nu, Bartalini:2011jp,Luszczak:2011zp,d'Enterria:2012qx,d'Enterria:2013ck,Maciula:2013kd}. 
The experimental evidence of the DPS from Tevatron and the LHC  has been presented  in 
\cite{Akesson:1986iv,Abe:1997bp,Abe:1997xk, Abazov:2009gc, Aad:2013bjm,Chatrchyan:2013xxa,Aad:2014rua}. 
At the LHC, the DPS is crucial for a better understanding of background for many important processes,  
e.g. the Higgs boson production \cite{DelFabbro:1999tf, Krasny:2013aca}, as well as for a better description of multiparton interactions needed for modeling the underlying event,  see  Refs.~\cite{Sjostrand:2004pf, Bartalini:2011jp}. It is, therefore,  very important to use a rigorous approach to the DPS which is based on QCD.

\begin{figure*}[t]
\centering\includegraphics[width = 6cm]{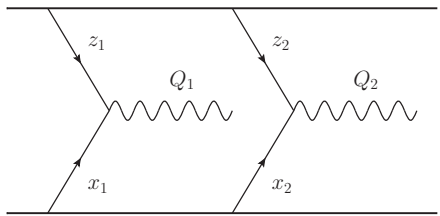}
\caption{Example of double parton scattering.}
\label{fig1}
\end{figure*}

The inclusive DPS cross section in the collinear approximation takes the form 
\cite{Diehl:2011yj}
\begin{eqnarray}\nonumber
\label{eq:cross1}
\sigma_{AB} &=& \frac{N}{2}
\sum_{f_1f_2f_1^\prime f_2^\prime}\int dx_1dx_2\,dz_1dz_2\, \frac{d^2{\bf q}}{(2\pi)^2}
\\ 
&\times& D_{f_1f_2} (x_1,x_2,Q_1,Q_2,{\bf q})\, {\hat{\sigma}}^A_{f_1f_1^\prime}(Q_1)\,
{\hat{\sigma}}^B_{f_2f_2^\prime}(Q_2)\, D_{f_1^\prime f_2^\prime}(z_1,z_2,Q_1,Q_2,-{\bf q}),
\end{eqnarray}
where  $A$ and $B$ denote two final states from two parton interactions with  hard scales
$Q_i=x_iz_i \sqrt{s}$, where $\sqrt{s}$ is center-of-mass energy of the colliding hadrons. In addition, 
$N$ is a symmetry factor, equal to 1 for $A=B$
and 2  otherwise,  and   $D_{f_1f_2}(x_1,x_2,Q_1,Q_2,{\bf q})$ are the {\it collinear} double parton distribution functions (DPDFs) in a hadron, see Fig.~\ref{fig1} for schematic illustration. They
 depend on parton longitudinal momentum fraction $x_{1,2}$ and  parton flavors $f_{1,2}$,    two
hard scales $Q_{1,2}$, and a relative transverse momentum ${\bf q}$. 
The latter momentum is related to the momentum structure of  four parton fields
in  the definition of unintegrated DPDFs, see \cite{Diehl:2011yj} for more details.  The  momentum fractions  obey the condition
\begin{eqnarray}
\label{eq:limit}
0< x_1+x_2 \le 1\,,
\end{eqnarray}
which means that the sum of parton longitudinal momenta cannot exceed the total nucleon momentum.
This is the basic parton correlation which has to be taken into account. For more advanced aspects of parton correlations see  Refs.~\cite{Diehl:2011yj,Manohar:2012jr}.

We start from the cross section formula in which the DPDFs depend on exchange momentum $\qbold$ rather than on the Fourier conjugate variable ${\bf b}$,
being interpreted as transverse distance between two partons taking part in hard scattering.  
The latter possibility as a starting point creates more problems than answers \cite{Diehl:2011tt,Diehl:2011yj}, despite its apparent attractiveness for phenomenological modeling of the DPDF dependence on this variable.

The DPDFs obey  QCD evolution equations known at present 
in the leading logarithmic approximation  \cite{Kirschner:1979im,Shelest:1982dg,Zinovev:1982be, Snigirev:2003cq,Korotkikh:2004bz}. These are 
the Dokshitzer-Gribov-Lipatov-Altarelli-Parisi (DGLAP)-type evolution equations with additional nonhomogeneous
terms which describe splitting of a single parton into two partons. The role of these terms for the DPS cross section
predictions is the main subject of this paper. We will focus on the electroweak boson production which is one of the cleanest processes for such an analysis. The electroweak bosons are color singlets, thus the collinear factorization   formula  (\ref{eq:cross1}) is not endangered  by  soft gluon final state interactions which might break factorization.

The paper is organized as follows. In Sec.\,\ref{section:2} we describe  evolution equations the DPDFs
in the leading logarithmic approximation. In Sec.~\ref{section:3} we derive the general solution to these equation
in the  Mellin moment space while in Sec.\,\ref{section:4} we present  assumptions concerning 
the relative momentum dependence of the DPDFs.
In Sec.\,\ref{section:5}  we apply the presented results to the computation of 
the electroweak boson production cross sections  and in Sec.\,\ref{section:6} we discuss the role of the 
contributions with the splitting terms. We conclude with  a summary of our findings.

\section{Evolution equations for DPDFs}
\label{section:2}

Evolution equations for the DPDFs 
are known in the leading logarithmic  approximation (LLA) in which large powers of $(\alpha_s\ln(Q^2/\Lambda^2))^n$ are resumed to all orders
in $n$. They were derived in \cite{Shelest:1982dg,Zinovev:1982be,Snigirev:2003cq,Korotkikh:2004bz} 
for equal hard scales, $Q_1=Q_2\equiv Q$, and 
for the relative  momentum $\qbold=0$,
\begin{eqnarray}\nonumber
\label{eq:twopdfeq} 
 \partial_t D_{f_1f_2}(x_1,x_2,t) &=&
 \sum_{f'}\int^{1-x_2}_{x_1} \frac{du}{u} \,{P}_{f_1f'}\!\left(\frac{x_1}{u}\right) D_{f' f_2}(u,x_2,t)
\\\nonumber
&+& \sum_{f'}\int_{x_2}^{1-x_1} \frac{du}{u}\,{P}_{f_2f'}\!\left(\frac{x_2}{u}\right)D_{f_1f'}(x_1,u,t)
\\\nonumber
\\
&+& \frac{1}{x_1+x_2} \sum_{f'}\,{P}_{f'\to f_1f_2}\!\left(\frac{x_1}{x_1+x_2}\right) D_{f'}(x_1+x_2,t)\,,
\end{eqnarray}
where we 
absorbed the leading order strong coupling constant into the definition of the evolution parameter,
\be\label{eq:tdef}
t\,\equiv\,t(Q)\,=\int^{Q^2}_{Q^2_0}\frac{\alpha_s(\mu^2)}{2\pi}\frac{d\mu^2}{\mu^2}\,, 
\ee
and introduced the shorthand notation for the DPDFs:
\be
D_{f_1f_2}(x_1,x_2,t) \equiv D_{f_1f_2}(x_1,x_2,Q,Q,\qbold=0)\,.
\ee
Notice that $t=0$ corresponds to an initial scale $Q_0$ at which the parton distributions need to be specified.
The first discussion of next-to-leading corrections to these equations can be found in 
\cite{Gaunt:2009re,Ceccopieri:2010kg}.
The integral kernels in Eq.~(\ref{eq:twopdfeq}) are the leading order 
Altarelli-Parisi splitting functions, with virtual corrections included, which describe 
the splitting of one of the two partons,  while the remaining parton stays intact.
This gives the upper integration limits resulting from condition (\ref{eq:limit}).

The last term on the rhs of Eq.~(\ref{eq:twopdfeq}), called from now on the {\it splitting term}, 
describes the real splitting of parton $f^{\prime}$ into two given partons $f_1$ and $f_2$.
The functions ${P}_{f\prime\to f_1f_2}$  are directly related to the real emission  
Altarelli-Parisi splitting functions in the LLA, $P_{f\prime f}^{(0)}$.  In particular, we have
\be
{P}_{q\to qg}(z) =  P_{qq}^{(0)}(z)\,,~~~~{P}_{q\to gq}(z) =  P_{qg}^{(0)}(z)\,,
~~~~{P}_{g\to q\qbar}(z) =  P_{gq}^{(0)}(z)\,,
~~~~{P}_{g\to gg}(z) =  P_{gg}^{(0)}(z)\,.
\ee 
The single parton distributions, $D_{f^\prime}(x_1+x_2,t)$, which appear in the splitting term,
provide an additional dependence on $t$ in Eq.~(\ref{eq:twopdfeq}), 
imposed by the DGLAP evolution equations, 
\be
\label{eq:onepdfeq}
\partial_t D_{f}(x,t) = \sum_{f^\prime}
\int_x^1\frac{du}{u}\,{P}_{ff^\prime}\!\left(\frac{x}{u}\right)D_{f'}(u,t).
\ee
The impact of the splitting terms on the DPS cross sections is the main subject of our paper.

\section{Solution to evolution equations}
\label{section:3}

Evolution equations (\ref{eq:twopdfeq})  greatly simplify in the space of
Mellin moments, obtained after the double Mellin transform of the DPDFs 
with respect to the momentum fractions 
$x_{1,2}$ which  obey condition (\ref{eq:limit}) imposed by the step function
$\Theta(1-x_1-x_2)$, 
\be\label{eq:doublemellin}
\Dtilde_{f_1f_2}(n_1,n_2,t) = \int_0^1dx_1\int_0^1dx_2\, x_1^{n_1}\, x_2^{n_1}\,
\Theta(1-x_1-x_2)\, D_{f_1f_2}(x_1,x_2,t).
\ee 
Introducing the matrix notation with respect to parton flavors (including gluon),
$\Dtilde=(\Dtilde_{f_1f_2})$, we find the new form of Eq.~(\ref{eq:twopdfeq}):
\begin{eqnarray}
\label{eq:twopdfeqmel} 
\partial_t \Dtilde(n_1,n_2,t) =
\gamma(n_1)\,\Dtilde(n_1,n_2,t) + \Dtilde(n_1,n_2,t)\,\gamma^T\!(n_2)\, +\, \gammatilde(n_1,n_2)\,\Dtilde(n_1+n_2,t)
\end{eqnarray}
where
\be\label{eq:anomdim}
\gamma(n) = \int_0^1 dx\,x^nP(x)\,,~~~~~~~~~~~~~
\gammatilde(n_1,n_2) = \int_0^1 dx\,x^{n_1}(1-x)^{n_2}P(x)
\ee
are known matrices of anomalous dimensions and $\Dtilde(n_1+n_2,t)$ is a vector
of the Mellin moments of the SPDFs,
\be
\Dtilde_f(n,t) = \int_0^1dx\,  x^{n}\, D_f(x,t)\,.
\ee 
They obey the DGLAP equation (\ref{eq:onepdfeq}) in the Mellin moment space,
\be 
\label{eq:onepdfeqmel}
\partial_t\Dtilde(n,t)\, =\, \gamma(n)\, \Dtilde(n,t)\,.
\ee

\begin{figure*}[t]
\centering\includegraphics[width = 9cm]{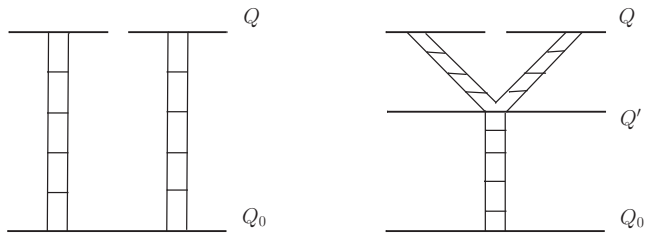}
\caption{Schematic illustration of the two contributions in solution (\ref{eq:solution1}).}
\label{fig2}
\end{figure*}

Equation~(\ref{eq:twopdfeqmel}) is a  nonhomogeneous first order linear differential  equation. Thus, its
solution is the sum of the general solution to
the homogeneous equation (without the splitting term) and a particular
solution to the nonhomogeneous equation. 
The homogeneous equation has the following general solution:
\be\label{eq:homsol}
\Dtilde(n_1,n_2,t) \,=\, {\rm e}^{\gamma(n_1)\,t}\,
A(n_1,n_2)\, {\rm e}^{\gamma^T\!(n_2)\,t}\,,
\ee
where the exponentials generate two DGLAP evolutions since the solution
to Eq.~(\ref{eq:onepdfeqmel}) reads
\be\label{eq:dglapsolution}
\Dtilde(n,t)\, =\,{\rm e}^{\gamma(n)\,t}\Dtilde_0(n)
\ee
where $\Dtilde_0(n)$ is an initial condition.
A particular solution to Eq.~(\ref{eq:twopdfeqmel} ) can now be found by making 
$A(n_1,n_2)$   time dependent.
Substituting such an ansatz to Eq.~(\ref{eq:homsol}), we find the equation
\be
\partial_t A(n_1,n_2,t) \,= \,
{\rm e}^{-\gamma(n_1)\, t}\,\gammatilde(n_1,n_2)\,\Dtilde(n_1+n_2,t)\,
{\rm e}^{-\gamma^T\!(n_2)\,t}
\ee
which can be easily solved:
\be\label{eq:A}
A(n_1,n_2,t) \,=\, \Dtilde_0(n_1,n_2)\, +\,
\int_{0}^{t} dt^\prime\,{\rm e}^{-\gamma(n_1)\,t^\prime}\,
\gammatilde(n_1,n_2)\,\Dtilde(n_1+n_2,t^\prime)\,
{\rm e}^{-\gamma^T\!(n_2)\,t^\prime}\,.
\ee
Thus, after substituting (\ref{eq:A})  in Eq.~(\ref{eq:homsol}), we obtain the final form of the solution 
to the evolution equations:
\be
\label{eq:solution1}
\Dtilde(n_1,n_2,t) = {\rm e}^{\gamma(n_1)\,t}\,
\Dtilde_0(n_1,n_2)\, {\rm e}^{\gamma^T\!(n_2)\,t}
\,+\,\int_{0}^{t} dt^\prime\,{\rm e}^{\gamma(n_1)(t-t^\prime)}\,
\gammatilde(n_1,n_2)\,\Dtilde(n_1+n_2,t^\prime)\,
{\rm e}^{\gamma^T\!(n_2)(t-t^\prime)}
\ee
where $\Dtilde_0(n_1,n_2)$ is an initial condition  at $t=0$. 
Solution (\ref{eq:solution1}) is the sum of two terms, see also Fig.~\ref{fig2}.
The first   term  describes two independent DGLAP evolutions  up to the scale $Q$ 
[related to $t$ by Eq.~(\ref{eq:tdef})] of two parton ladders emerging from a hadron 
at the initial  scale $Q_0$ ($t=0$).  The  second term describes the emergence of two parton ladders from
a single parton ladder through the splitting at the scale $Q^\prime$ 
(corresponding to  $t^\prime$) and  their  independently evolution up to the scale $Q$.
Solution  (\ref{eq:solution1}) can also be written in the $x$ space using the inverse Mellin transform, see Ref.~\cite{Snigirev:2011zz}.

Notice that the first term in Eq.~(\ref{eq:solution1}) is a solution to the homogeneous equation (\ref{eq:twopdfeqmel}) without the splitting term, $\Dtilde^{(hom)}$,
which depends on the initial condition for the DPDFs, while
the second term, which we denote by $\Dtilde^{(nhom)}$, depends only on the initial condition for the SPDFs.  Thus, it  can be computed  for any initial conditions for DPDFs as the difference between
the solutions to the nonhomogeneous and the homogeneous equations.
It can also  be directly obtained by solving to the nonhomogeneous equation 
with zero initial conditions for DPDFs. 

\begin{figure*}[h]
\centering\includegraphics[width = 15cm]{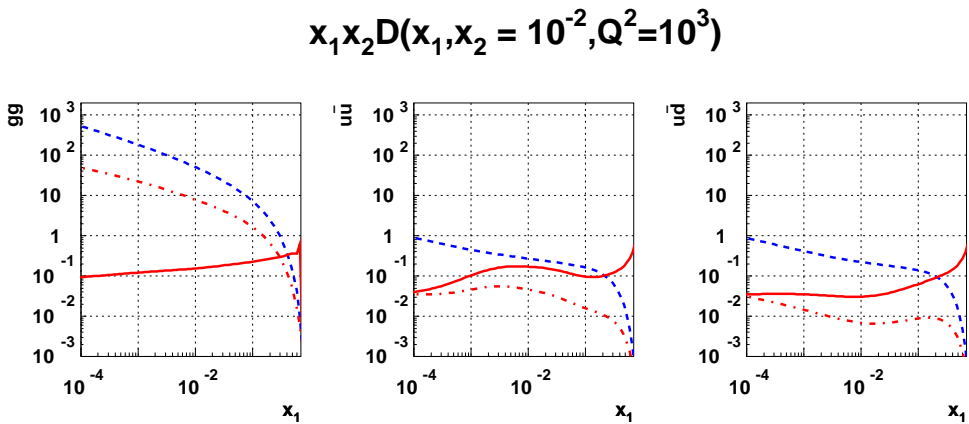}
\caption{The homogeneous, $\Dtilde^{(hom)}$ (dashed lines),  and nonhomogeneous, $\Dtilde^{(nhom)}$
(dash-dotted lines), solutions for the indicated flavors. The ratios of the two solutions are shown as the solid lines.}
\label{fig2a}
\end{figure*}

In Fig.~\ref{fig2a} we plot the DPDFs for the indicated flavors
as functions of $x_1$ for fixed $x_2=10^{-2}$ and $Q^2=10^3~{\rm GeV}^2$. We see that for $x<0.1$, 
the splitting part of the solution, $\Dtilde^{(nhom)}$, is significantly smaller than $\Dtilde^{(hom)}$
with the ratio of the order of  $10^{-1}$. 

The curves in Fig.~\ref{fig2a}, as well as the rest of the numerical results shown in this paper, were obtained
with our numerical program \cite{GBeq} which solves evolution equations 
(\ref{eq:twopdfeq}) and (\ref{eq:onepdfeq}), 
using the Chebyshev polynomial expansion on the grid of Chebyshev nodes,  
$y_i\in[-1,1]$, see \cite{Press:1992zz} for more details. These nodes were subsequently transformed
into the range appropriate for the parton distribution functions, $x\in [x_{\rm min},1]$,  through the logarithmic transformation $y=A\ln x +B$.  
We used $N=30$ nodes  for  $x_{\rm min}=10^{-5}$, and  10 point grid for the  variable $t$
in the range corresponding to $Q\in [2,200]~{\rm GeV}$.

\section{Relative momentum dependence}
\label{section:4}

The form (\ref{eq:solution1})  of the solution is the basis of the proposition of Ryskin and Snigirev 
 \cite{Ryskin:2011kk} concerning  the dependence of the DPDFs on the  relative momentum $\qbold$. 
Such a dependence  is not specified by the evolution equations
and is a matter of a physically motivated modeling. The basic idea is that 
for two partons  originating from a nucleon,  $\qbold$ 
reflects their correlation  inside the nucleon described by a 
nonperturbative form factor.  On the other hand, if the two partons originate from a pointlike parton through its splitting,  the form factor no longer exists. 

Therefore, 
the first term in  Eq.~(\ref{eq:solution1})  has been postulated in \cite{Ryskin:2011kk} 
with the factorized $\qbold$ dependence,
\be\label{eq:solution2a}
\Dtilde^{(1)}(n_1,n_2,t,\qbold) \,=\, {\rm e}^{\gamma(n_1)\,t}\,
\Dtilde_0(n_1,n_2)\, {\rm e}^{\gamma^T\!(n_2)\,t}\,F_{2g}^2(\qbold)
\ee
where $F_{2g}(\qbold)$ is the two-gluon nucleon form factor in the dipole form
\be\label{eq:formfactor}
F_{2g}(\qbold)\,=\,\frac{1}{\left(1+{\qbold^2}/{m_g^2}\right)^2}
\ee
and $m_g$ is the effective gluon mass. In principle, the form factor could depend on parton flavors, 
however, this dependence is not taken into account. 

In the second term in Eq.~(\ref{eq:solution1}), the $\qbold$ dependence has been introduced  through
the lower integration limit,
\be\label{eq:solution2b}
\Dtilde^{(2)}(n_1,n_2,t,\qbold)\,=\,\int_{t_0}^{t} dt^\prime\,{\rm e}^{\gamma(n_1)(t-t^\prime)}\,
\gammatilde(n_1,n_2)\,\Dtilde(n_1+n_2,t^\prime)\,
{\rm e}^{\gamma^T\!(n_2)(t-t^\prime)}
\ee
where 
\be\label{eq:lowerlimit}
t_0\,=\,\left\{ \begin{array}{cl}
 t(|\qbold|) & \mbox{if~~$Q_0<|\qbold| \le Q$}
  \\ \\
 0 &\mbox{if~~$|\qbold|\le Q_0$} 
       \end{array} \right.
\ee
and $t(|\qbold|)$ is given by Eq.~(\ref{eq:tdef}). Thus, for $|\qbold|> Q_0$, $|\qbold|$ is the scale from which the splitting starts.
For $|\qbold|< Q_0$, the relative loop momentum is small 
and may be neglected due to strong ordering in transverse parton momenta in the DGLAP ladder.
In such a case, the integration in Eq.~(\ref{eq:solution2b}) starts from $Q_0$ which corresponds 
to $t_0=0$. The values of $|\qbold|$ were restricted to $|\qbold|\le Q$,   which means that $Q$ is the largest scale in the problem.

The two components  given by
Eqs.~(\ref{eq:solution2a}) and (\ref{eq:solution2b}) can also be written in 
the $x$ space, see \cite{Ryskin:2011kk}. In this way, the general form of the DPDFs
reads
\be\label{eq:solution22}
D(x_1,x_2,Q,\qbold) \,=\,D^{(1)}(x_1,x_2,Q,\qbold) \,+\,D^{(2)}(x_1,x_2,Q,\qbold)\,,
\ee
where we reintroduced  the hard scale $Q$ in the notation [corresponding to $t=t(Q)$].
Cross section (\ref{eq:cross1}) can be written in terms of these components  as
the  sum 
\be\label{eq:cross11}
\sigma_{AB} \,=\,\sigma_{AB}^{(11)}\,+\,\sigma_{AB}^{(12+21)}+\,\sigma_{AB}^{(22)}
\ee
where
\be\nonumber
\label{eq:cross11a}
\sigma_{AB}^{(ij)}\!&=&\!\frac{N}{2}
\sum_{f_i,f_i^\prime}\int dx_1dx_2\,dz_1dz_2\, \int\frac{d^2{\bf q}}{(2\pi)^2}\,
\theta(Q-|\qbold|)
\\ 
&\times&\! D_{f_1f_2}^{(i)}(x_1,x_2,Q,{\bf q})\, {\hat{\sigma}}^A_{f_1f_1^\prime}(Q)\,
{\hat{\sigma}}^B_{f_2f_2^\prime}(Q)\, 
D_{f_1^\prime f_2^\prime}^{(j)}(z_1,z_2,Q,-{\bf q}).
\ee
Notice that the integration over $\qbold$ is bounded from above by the hard scale $Q$.
Each term in the above sum has a clear interpretation;  $\sigma_{AB}^{(11)}$ is a contribution without parton splitting, 
$\sigma_{AB}^{(12+21)}$ is a single splitting contribution while
$\sigma_{AB}^{(22)}$ is a double splitting contribution with two parton splittings from both hadrons each.
The latter contribution was a matter of intensive debate in past years, see 
\cite{Diehl:2011tt,Gaunt:2011xd,Blok:2011bu,Diehl:2011yj,Manohar:2012pe,Gaunt:2012dd},
with a conclusion that it  should rather be classified as the single parton scattering process
since it is entirely driven by the SPDFs. However, it was advocated in Ref.~\cite{Gaunt:2012dd} that the complete
removal of the double splitting graphs from the DPS cross section is not the quite correct prescription. 
All this means that the double splitting contribution needs careful diagrammatic analysis. 
Thus, we do not consider the $\sigma_{AB}^{(22)}$ contribution in our forthcoming presentation, leaving
the problem of the double splitting graphs to a separate analysis.

In the standard approach, the estimation of  DPS cross sections is usually made with the formula
\be\label{eq:basicdps}
\sigma_{AB} \,=\,\frac{N}{2}\,\frac{\sigma_A\sigma_B}{\sigma_{\rm eff}}\,,
\ee
where $\sigma_{A}$ and $\sigma_{B}$  are the single parton scattering cross sections 
and $\sigma_{\rm eff}$ is an effective cross section, present here for the dimensional reason. 
The CDF and D0 collaborations estimated the value $\sigma_{\rm eff}\approx 15~{\rm mb} $
from the DPS dijet data \cite{Abe:1997bp,Abe:1997xk,Abazov:2009gc}.  Comparing (\ref{eq:basicdps})
to the standard contribution $\sigma_{AB}^{(11)}$ we see that 
$\sigma_{\rm eff}$ is the inverse of the integral,
\be
\label{eq:formfactor1}
\int \frac{d^2{\bf q}}{(2\pi)^2}\,\theta(Q-|\qbold |)\,F^4_{2g}(\qbold)=\frac{m_g^2}{28\pi}
\ee
for $Q\gg m_g$, which leads to effective gluon mass  $m_g\approx 1.5~{\rm GeV}$. 

\begin{figure*}[t]
\centering\includegraphics[width = 6cm]{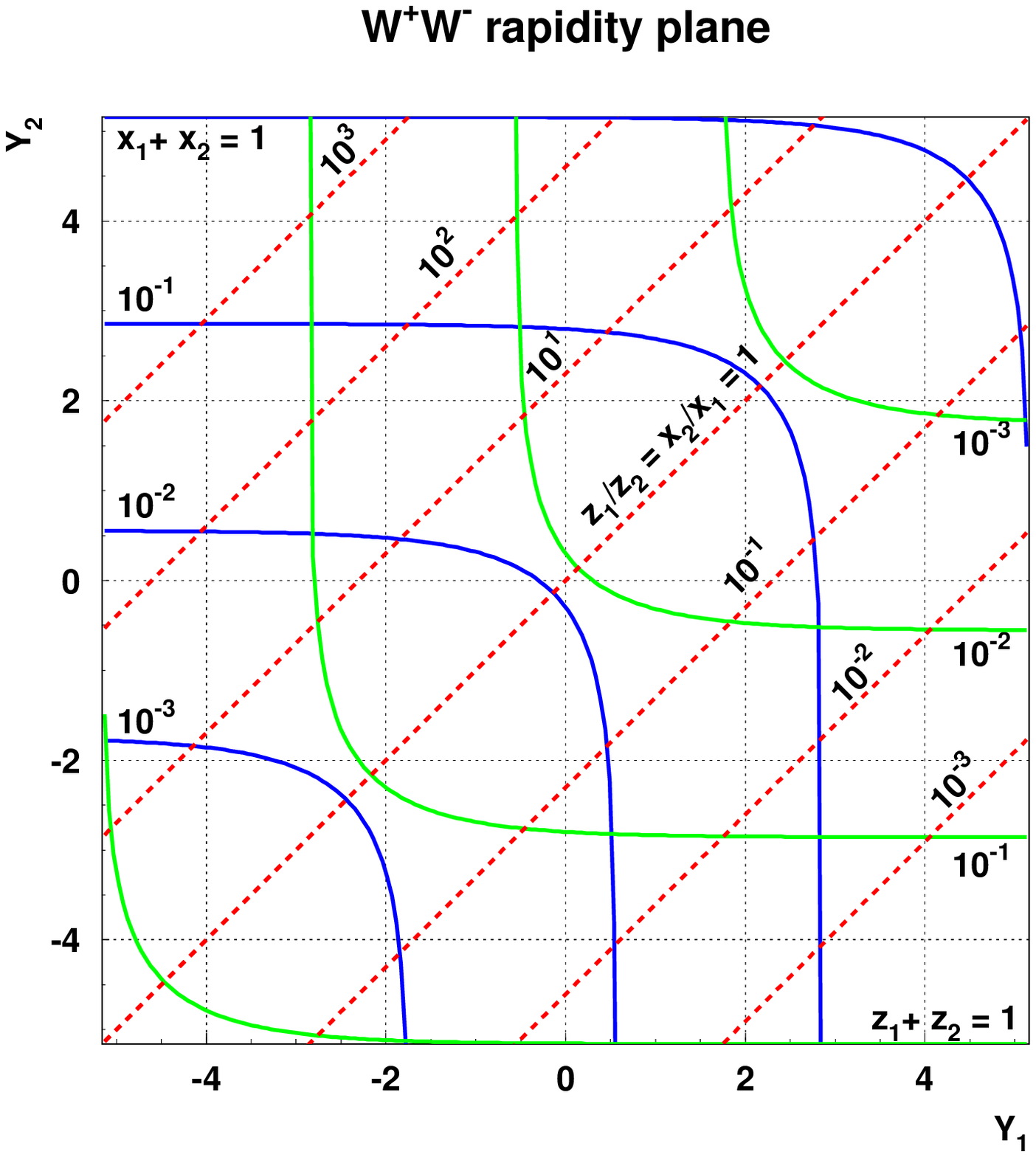}
\centering\includegraphics[width = 6cm]{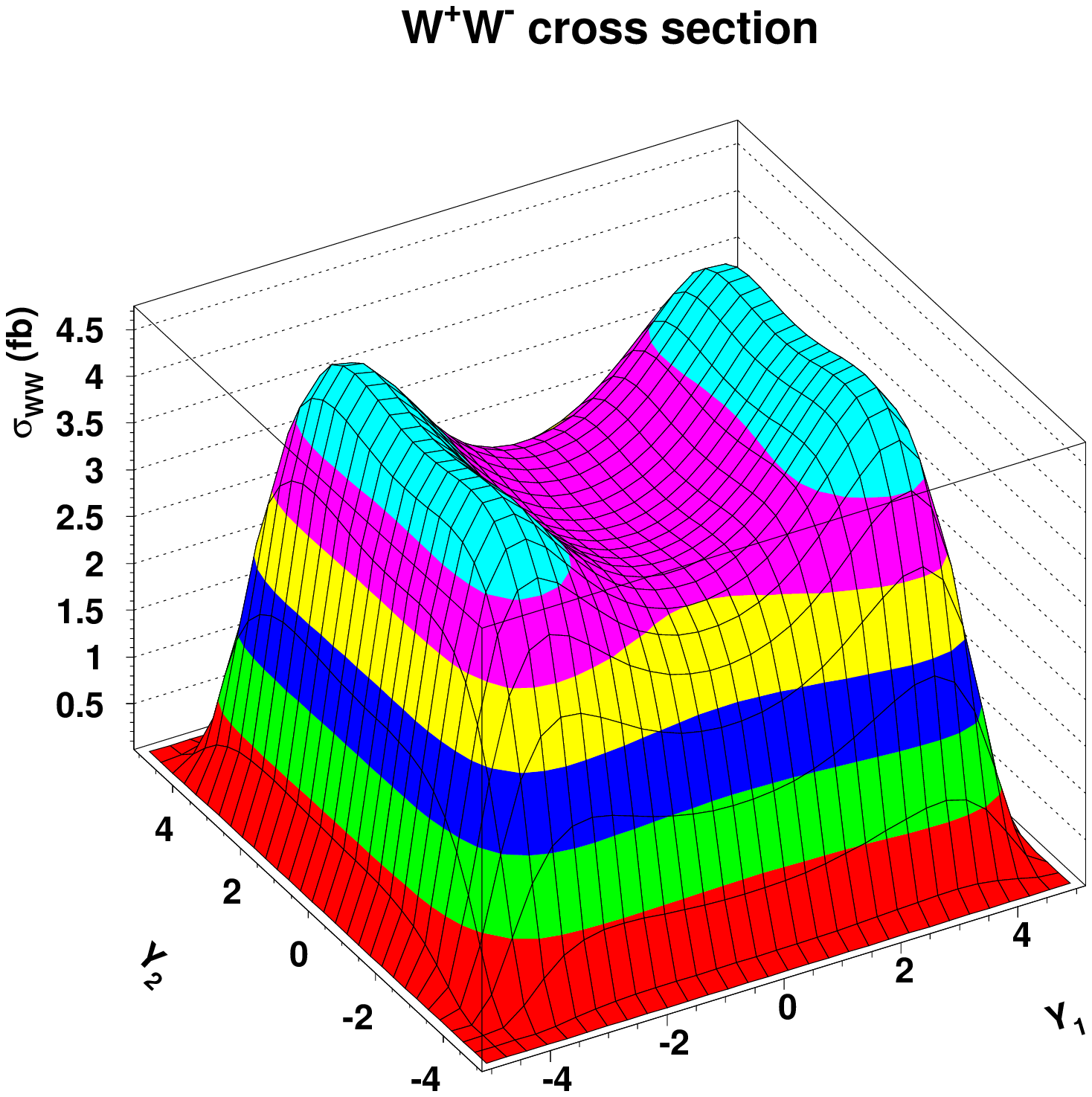}
\caption{Rapidity plane for $W^+W^-$ boson production (left)  and cross section  
(\ref{eq:csfact}) (in femtobarns) (right).}
\label{fig5}
\end{figure*}

\section{Electroweak boson production in DPS}
\label{section:5}

\begin{figure*}[t]
\centering\includegraphics[width = 15cm]{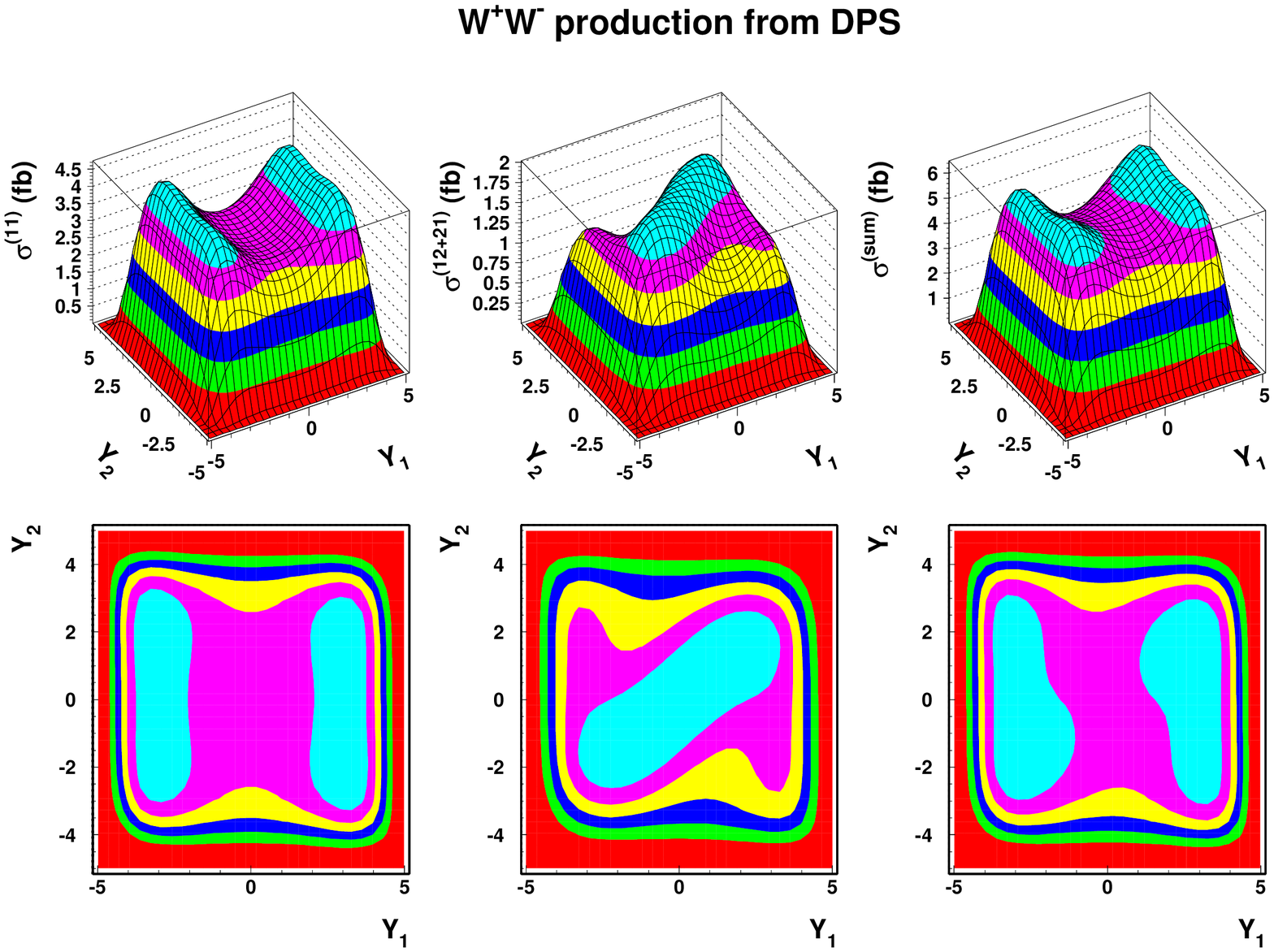}
\caption{The two contributions to $W^+W^-$ production cross section (\ref{eq:cross11})  together with their sum
(in femtobarns), differential in boson rapidities $y_{1,2}$. The contours of constant values are shown below.}
\label{fig6}
\end{figure*}

As an application of the presented formalism, we consider the DPS electroweak
boson production $W^+W^-$ and $Z^0Z^0$ in the proton-proton scattering at the LHC center-of-mass energy 
$\sqrt{s}=14~{\rm TeV}$.  The  hard scale in this case is given by the boson
mass, $Q=M_{W,Z}$. We compute the DPS cross section (\ref{eq:cross11a}), $d^2\sigma_{AB}/dy_1dy_2$, differential in boson rapidities $y_{1,2}$. In such a case, in the collinear approach, the parton momentum fractions obey the condition
\be
x_{1,2}=\frac{Q}{\sqrt{s}}\,\eto^{y_{1,2}}\,,
~~~~~z_{1,2}=\frac{Q}{\sqrt{s}}\,\eto^{-y_{1,2}}\,.
\ee
The allowed values of rapidities, resulting
from the relations $x_{1,2},z_{1,2},(x_{1}+x_{2}),(z_{1}+z_{2})\in [0,1]$, 
are shown in Fig.~\ref{fig5} (left). The solid lines correspond to
 constant values of $(x_1+x_2)$  and $(z_1+z_2)$, while
the dashed lines denote constant ratios $x_2/x_1=z_1/z_2$.

In Fig.~\ref{fig5} (right) we also show the DPS cross section computed using the formula
(\ref{eq:basicdps}) with factorized hard interactions
\be
\frac{d^2\sigma_{W^+W^-}}{dy_1dy_2}\,=\,\frac{1}{\sigma_{\rm eff}}\frac{d\sigma_{W^+}}{dy_1}
\frac{d\sigma_{W^-}}{dy_2}
\label{eq:csfact}
\ee
where $\sigma_{\rm eff}\approx 15~{\rm mb}$. The single scattering cross sections read
\be
\frac{d\sigma_{W^{\pm}}}{dy} = \sigma_0^W\sum_{qq^\prime}|V_{qq^\prime}|^2
\left\{q(x_+,M_W)\,\overline{q}^\prime(x_-,M_W)\,+\,\overline{q}(x_+,M_W)\,q^\prime(x_-,M_W)
\right\},
\ee
where $q,\overline{q}$ are the appropriate quark/antiquark distributions, 
$V_{qq^\prime}$ is the Kobayashi-Maskawa matrix and
\be
\sigma_0^W=\frac{2\pi G_F}{3\sqrt{2}}\frac{M^2_W}{s}\,,~~~~~~~~~~~~~~~
x_\pm = \frac{M_W}{\sqrt{s}}\,\eto^{\pm y}\,.
\ee
We used three quark flavors in the computations and the leading order MSTW08 parametrization of
the SPDFs \cite{Martin:2009iq}.

\begin{figure*}[t]
\centering\includegraphics[width = 15cm]{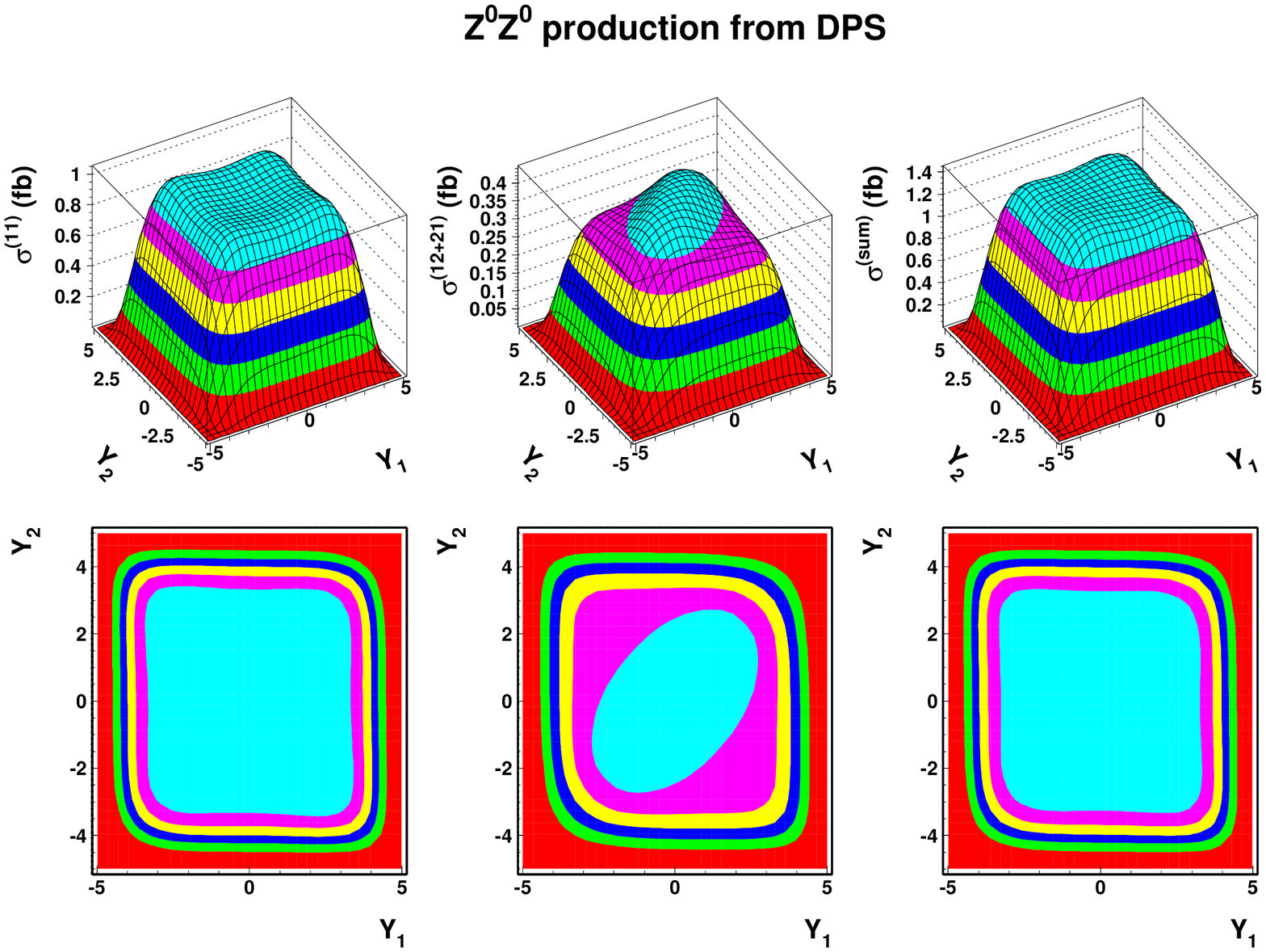}
\caption{The same as in Fig.~\ref{fig6} but for $Z^0Z^0$ bosons.}
\label{fig7}
\end{figure*}

To compute the DPS cross sections (\ref{eq:cross11}),  we performed  decomposition (\ref{eq:solution22}) 
with our numerical program, following the method described in Sec.~\ref{section:3}. We found
the solution to the homogeneous evolution equations  with the initial condition suggested in 
Ref.~\cite{Gaunt:2009re}, $D^{(hom)}$, and the solution to the nonhomogeneous equations with zero initial conditions,
$D^{(nhom)}$.  The two components in Eq.~(\ref{eq:solution22})  are written with the help of the found solutions
\be
\label{eq:d1}
D^{(1)}(x_1,x_2,Q,\qbold) &=& D^{(hom)}(x_1,x_2,Q)F^2_{2g}(\qbold)
\\\nonumber
\\\label{eq:d2}
D^{(2)}(x_1,x_2,Q,\qbold) &=& D^{(nhom)}(x_1,x_2,Q)- D^{(nhom)}(x_1,x_2,|\qbold|)
\ee
where the subtraction in Eq.~(\ref{eq:d2}) accounts for  the lower integration limit  in Eq.~(\ref{eq:solution2b}).

The two contributions to the cross section $d\sigma_{AB}/dy_1dy_2$ 
are shown  for the $W^+W^-$ production  in Fig.~\ref{fig6}
and  for the $Z^0Z^0$ production in Fig.~\ref{fig7}.   
We see that in both cases the single splitting contribution, $\sigma^{(12+21)}$, 
is comparable with  the standard contribution, $\sigma^{(11)}$.  Notice also that the latter contribution stays very close to the factorized form
(\ref{eq:csfact}), see  Fig.~\ref{fig5} (right). We quantify these observations in the next section.
In Table I we present the values of these contributions to the total cross sections,
obtained after  the integration over the allowed values of boson rapidities.
\vskip 0.3cm

\begin{table}[h]
\label{table1}
\begin{tabular}{|c|c|c|c|}
\hline
    in [fb]              &~~~$\sigma_{tot}^{(11)}~~~$  &~~$\sigma_{tot}^{(12+21)}$~~&~  $\sigma_{tot}^{(12+21)}/\sigma_{tot}^{(11)}$~ \\
                  \hline
~~$W^+W^-$~~ &   256       &     97      & 0.38    \\
                  \hline
~$Z^0Z^0$~~~ &     61    &     22     &  0.36 \\
\hline
\end{tabular}
\caption{Contributions to the total DPS cross sections for electroweak boson production.}
 \end{table}

Some remarks concerning the correlation pattern in rapidities are in order, especially for the $W^+W^-$ production.
The standard contribution, $\sigma^{(11)}$, is dominated by the production of the $W^+$ boson with rapidity $y_1\approx \pm 4$
and the $W^-$ boson in the broad range of the rapidity  $y_2$. 
This is well understood from the factorized form (\ref{eq:csfact}) of the DPS cross section, given in terms of the SPS cross sections
with the characteristic shapes in rapidity, see e.g.  Ref.~\cite{GolecBiernat:2009pj} for these shapes in the LHC kinematics.  
Such a pattern reflects the dominant mechanism of the $W^+$ production from the valence $u$ quark with the longitudinal momentum fraction
$x\approx 0.3$ which annihilates with the sea $\dbar$ quark. On the other hand, in the single splitting contribution, $\sigma^{(12+21)}$,
the $W^\pm$ bosons are produced mostly with rapidities which are correlated along the line of equal rapidities, $y_1=y_2$.  The sum of 
the two contributions in Fig.~\ref{fig6} shows that the single splitting contribution leads to the distortion of the rapidity
correlation pattern in comparison to the standard contribution correlation. We hope that this could be measured at the LHC.

\section{Discussion of the splitting contribution}
\label{section:6}

\begin{figure*}[t]
\centering\includegraphics[width = 7cm]{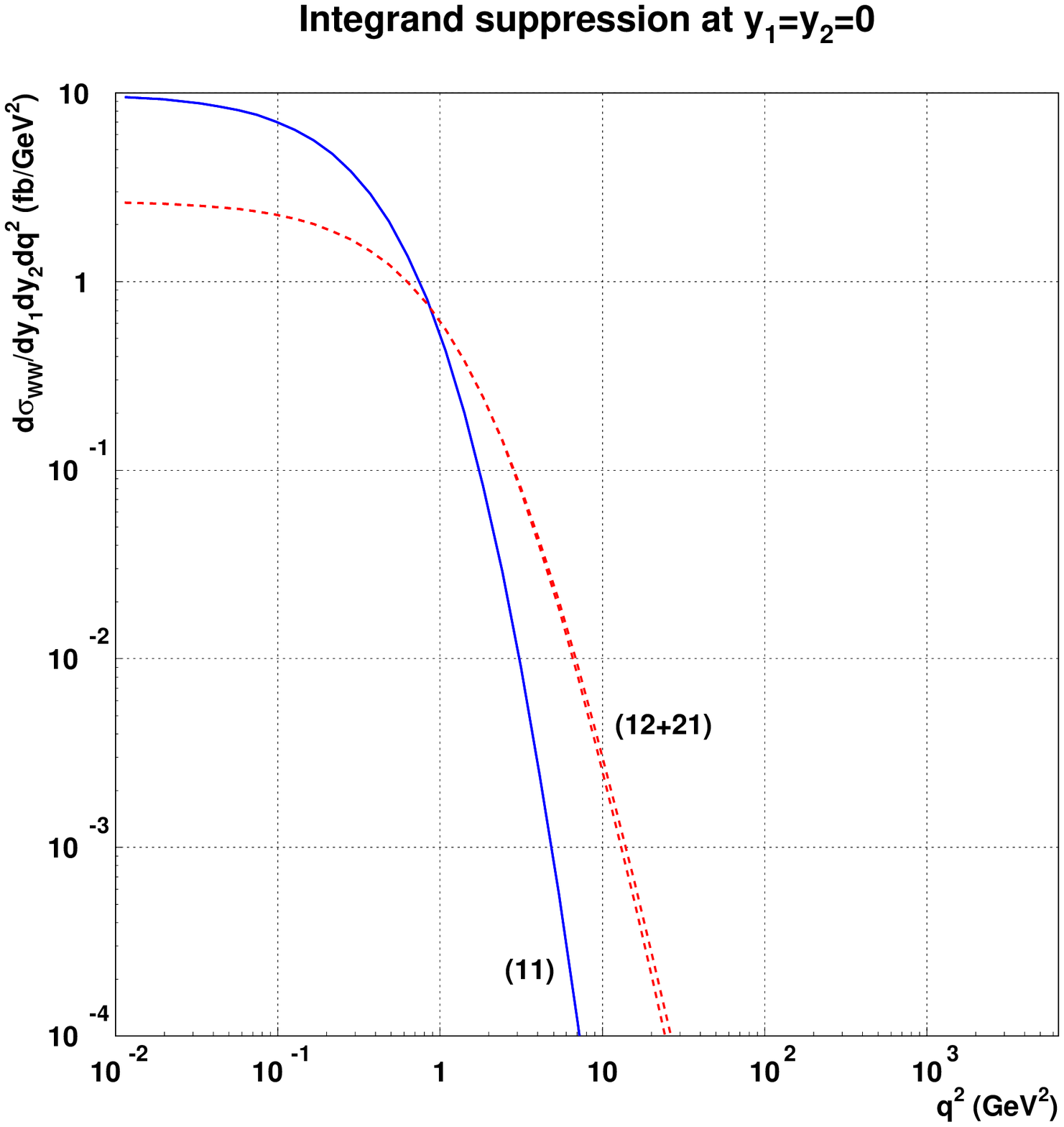}
\caption{The $\qbold^2$ dependence of $d\sigma_{W^+W^-}/dy_1dy_2d\/\qbold^2$ for the indicated contributions.
The upper limit for $\qbold^2$ equals $M_W^2$.}
\label{fig8}
\end{figure*}

To understand to the origin of the ratios in Table I, we plot
the cross section  $d\sigma_{W^+W^-}/dy_1dy_2 d\/\qbold^2$ as a function of $\qbold^2$ for 
the indicated  contributions taken at the central rapidities, $y_1=y_2=0$, see Fig.~\ref{fig8}.
As expected, the $(11)$ and $(12+21)$ contributions are suppressed for large values of $q^2$
because of the presence of the form factor $F_{2g}(\qbold)$ to the power 4 and 2, respectively, in these contributions.
We found that the dependence on $\qbold$ of  the nonhomogeneous distribution $D^{(2)}$   in the contribution $(12+21)$, 
given Eq.~(\ref{eq:solution2b}),  is negligible, which 
is shown by the two dashed lines in Fig.~\ref{fig8}.
Thus,  the   single splitting  contribution integrated over $\qbold$  is  proportional to the integral 
\be
\label{eq:formfactor2}
\int \frac{d^2{\bf q}}{(2\pi)^2}\,\theta(Q-|\qbold|)\,F^2_{2g}(\qbold)\,=\,\frac{m_g^2}{12\pi}\,.
\ee
In this way, we find the following ratios from the $\qbold$ dependence of the two contributions
for $m_g=1.5~{\rm GeV}^2$,
\be\label{eq:ratio1}
\frac{m_g^2}{28\pi} : \frac{m_g^2}{12\pi} = 1 : 2.33\,.
\ee
The significant enhancement of the single splitting contribution due to the weaker $q^2$-dependence 
caused to the presence of the  nonhomogeneous component, 
$D^{(2)}$, is compensated by its smaller size in comparison to the homogeneous component, $D^{(1)}$. 
Roughly speaking, in $\sigma^{(11)}$ the DPDFs are proportional to $(x^{-\lambda})^4$ with $\lambda \sim 0.3-0.5$ at $x<0.1$ while in 
$\sigma^{(12+21)}$ the  DPDFs are only proportional to $(x^{-\lambda})^3$. More precisely, 
the  ratio of the DPDFs taken for $y_1=y_2=0$ in the two contributions  can be found from the values of the cross sections 
at $\qbold^2\approx 0$ in Fig.~\ref{fig8},
\be\label{eq:ratio2}
DPDF^{(11)}:DPDF^{(12+21)} = 1 : 0.27\,,
\ee
which is in reasonable agreement with the results shown in Fig.~\ref{fig2a}. 
Multiplying the ratios (\ref{eq:ratio1}) and (\ref{eq:ratio2}) we find the ratio of the differential cross sections at 
$y_1=y_2=0$, which can be read off from  Fig.~\ref{fig6}:
\be\label{eq:csratios}
\sigma^{(11)} : \sigma^{(12+21)} = 1 : 0.63\,.
\ee
This ratio is  bigger than those for the total cross sections in Table I. 
Nevertheless, the mechanism explaining these ratios  is all  the same.

\begin{figure*}[t]
\centering\includegraphics[width = 12cm]{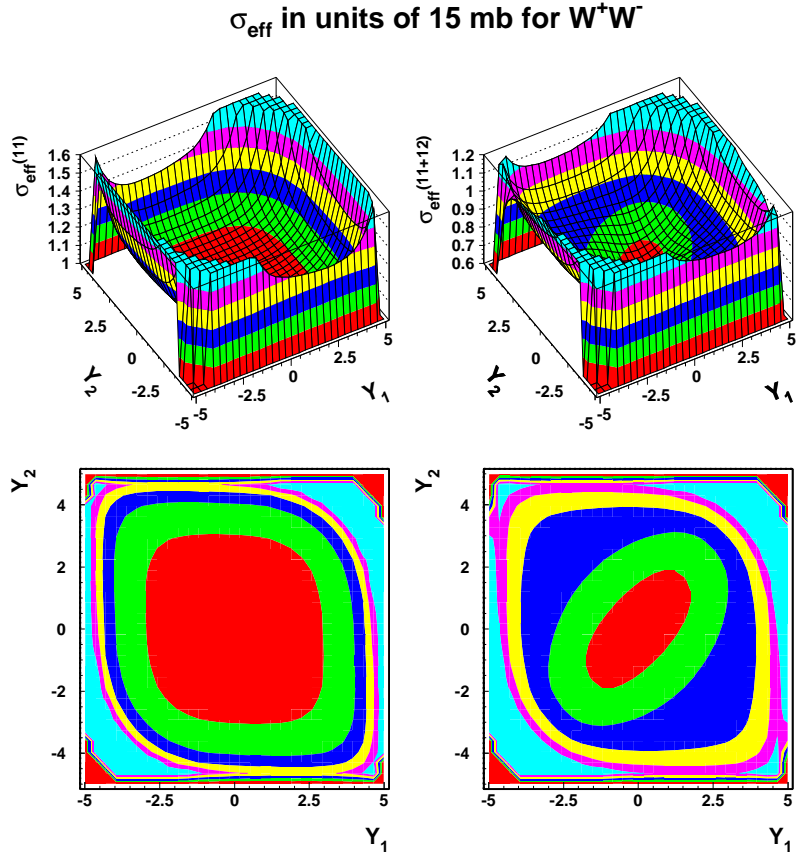}
\caption{$\sigma_{\rm eff}$ 
in units of 15 mb for $W^+W^-$ production as a function of $(y_1,y_2)$ for 
$\sigma_{AB}=\sigma_{AB}^{(11)}$  (left) and  $\sigma_{AB}=\sigma_{AB}^{(11)}+\sigma_{AB}^{(12+21)}$ (right).  
The lines of constant values are shown below.}
\label{fig9}
\end{figure*}

From the presented results we confirm the observation of Ref.~\cite{Blok:2011bu} that  
the one splitting contribution should be considered in all analyses. 
We will show its significance
for the estimation of the effective cross section for the electroweak boson production.
Following Eq.~(\ref{eq:basicdps}), we define
\be
\sigma_{\rm eff} = \frac{N}{2}\frac{(d\sigma_A/dy_1)(d\sigma_B/dy_2)}{d\sigma_{AB}/dy_1dy_2}
\ee
for the two cases: $\sigma_{AB}=\sigma_{AB}^{(11)}$ and 
$\sigma_{AB}=\sigma_{AB}^{(11)}+\sigma_{AB}^{(12+21)}$. Obviously, $\sigma_{\rm eff}$ will depend
on boson rapidities $(y_1,y_2)$, which dependence illustrates  the violation of a simple-minded assumption that
the DPS cross section is proportional to the product of the SPS cross sections. 

\begin{figure*}[t]
\centering\includegraphics[width = 12cm]{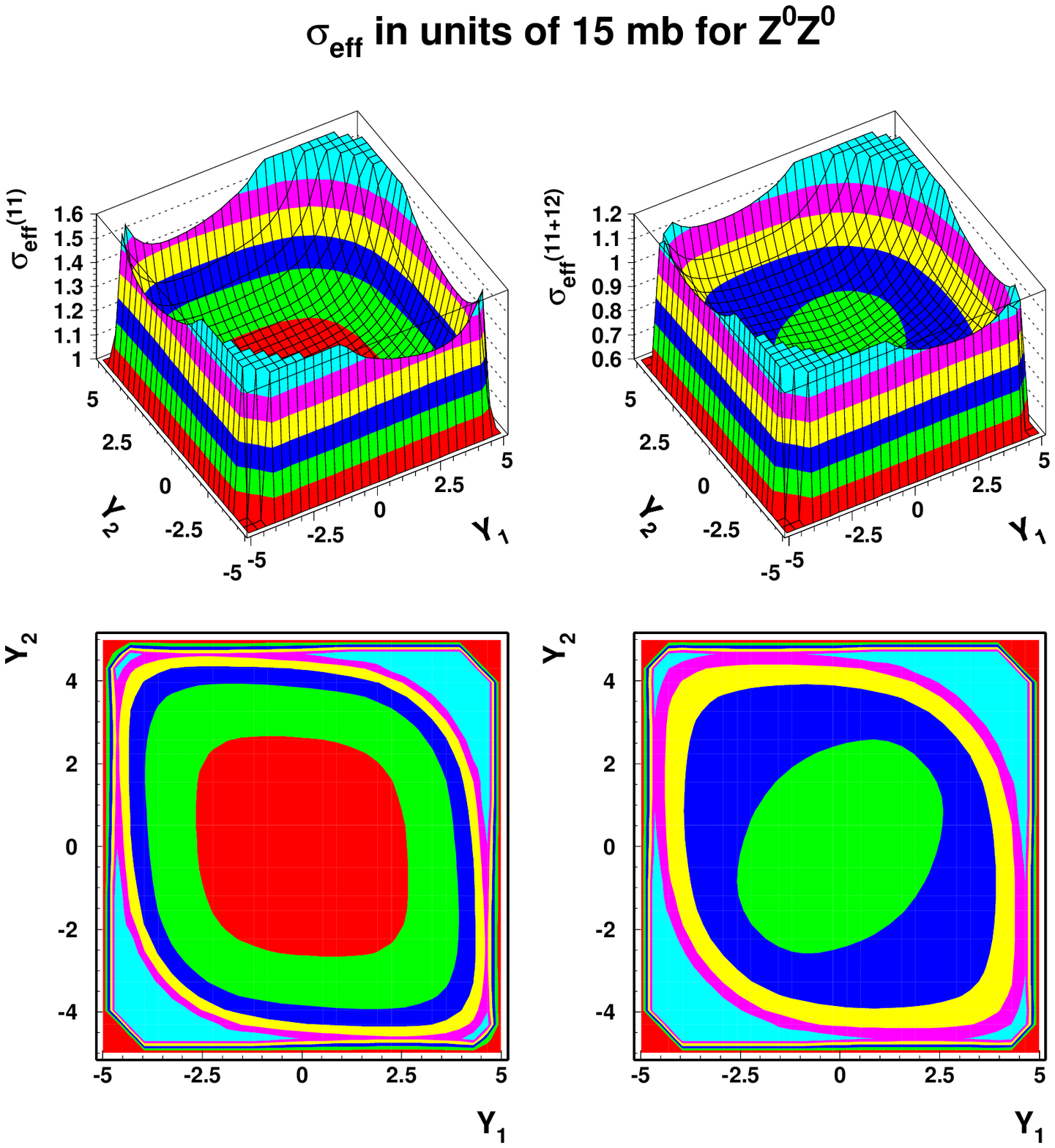}
\caption{The same as in Fig.~\ref{fig9} but for $Z^0Z^0$ production.}
\label{fig10}
\end{figure*}

In Figs.~\ref{fig9} and \ref{fig10} we show $\sigma_{\rm eff}$ for the $W^+W^-$ and $Z^0Z^0$ production,
respectively, in the two cases specified above, in units of the standard value of the effective cross section,
$\sigma_{\rm eff}\simeq 15~{\rm mb}$. For better visibility, we cut the maximal values to 1.6 or 1.2 at the edges of the phase space.
We see that with the standard contribution to the DPS cross section, the
factorization property is to good approximation valid in the central region of rapidities (small
values of parton momentum fractions, see Fig.~\ref{fig5} for the correspondence). However, approaching kinematic boundaries $x_1+x_2=z_1+z_2=1$ with comparable momentum  fractions, 
the violation of factorization becomes stronger. This picture changes after adding the one splitting contribution. Now,
the violation of factorization is significant even in the central rapidity region. The effective cross
section is smaller than $15~{\rm mb}$,  in between $60\%$-$80\%$ of this value ($9$-$12$~mb).

\section{Summary}

We have analyzed the DPS processes in the collinear approximation, using evolution equations (\ref{eq:twopdfeq}) for the DPDFs.  We have  concentrated on the significance of the splitting terms in these equations
for the DPS processes with a large hard scale $Q\sim 100~{\rm GeV}$. For the illustration, 
we considered $W^+W^-$ and
$Z^0Z^0$ boson production at the LHC center-of-mass energy ${14}~{\rm TeV}$. To compute the DPS cross sections, 
we have specified the dependence of the DPDFs on the relative momentum $\qbold$, following Ref.~\cite{Ryskin:2011kk}. 
In this model, the splitting component of the DPDFs is not strongly suppressed  
at large values of the relative momentum fraction $|\qbold|$, like the standard component, because it originates from the splitting  
of a pointlike parton. 
Based on the  constructed numerical program which solves the evolution equations for DPDFs, 
we analyzed  the single splitting contribution to the DPS cross sections for the electroweak boson production.
We  quantified  the relevance of the single splitting  contribution in such a case in terms of the effective cross section.
We also discussed correlations in rapidity for the produced $W^\pm$ bosons pointing out that the single splitting contribution
distorts the standard correlation obtained with the factorized DPS cross section. 

\bigskip
{\it Note added in proof} - Once the first version of this  paper was released we found that Ref.~\cite{Gaunt:2014rua} appeared
which addresses the same problems but for different DPS cross sections, using the same methods.

\begin{acknowledgments}
This work was supported by the  Polish NCN Grant No. DEC-2011/01/B/ST2/03915  and by the Center
for Innovation and Transfer of Natural Sciences and Engineering Knowledge in Rzesz\'ow.

\end{acknowledgments}

\section{REFERENCES}
\bibliographystyle{h-physrev4}
\bibliography{mybib}

\end{document}